\documentstyle[aps,prb,epsfig,multicol]{revtex}
\begin{document}

\draft
%\preprint{SNUTP 00-XXX}

\title{Renormalization-group study of gate charge effects in
Josephson-junction chains}
\author{M.Y. Choi$^{1,2}$, Sung Wu Rhee$^2$, Minchul Lee$^1$, and J. Choi$^3$} 
\address{$^1$Department of Physics, Seoul National University,
Seoul 151-742, Korea}
\address{$^2$Department of Physics, University of Washington,
Seattle WA 98195, U.S.A.}
\address{$^3$Department of Physics, Keimyung University,
Taegu 704-701, Korea}

%\date{Version 0.4}
\maketitle

\begin{abstract}
We study the quantum phase transition in a chain of superconducting grains, 
coupled by Josephson junctions, with emphasis on the effects of
gate charges induced on the grains.
At zero temperature the system is mapped onto a two-dimensional classical
Coulomb gas, where the gate charge plays the role of an imaginary 
electric field.
Such a field is found relevant in the 
renormalization-group transformation and to change the nature of 
the superconductor-insulator transition present in the system,
tending to suppress quantum fluctuations and 
helping establish superconductivity.
On the basis of this observation,
we propose the zero-temperature phase diagram on the plane of
the gate charge and the energy ratio.
\end{abstract}

\bigskip


\begin{multicols}{2}

In arrays of ultra-small superconducting grains,
the charging energy can be dominant over the
Josephson coupling energy, giving rise to crucial effects of 
quantum fluctuations.~\cite{Schonx90}  In particular recent advances in
fabrication techniques make it possible to control the physical 
parameters of the arrays, providing a convenient model system
for the study of quantum phase transitions~\cite{Sondhi97}
between superconducting and insulating phases.
In such an array of ultra-small junctions,
frustration can be introduced not only by applying magnetic fields
but also by inducing gate charges; these control the number of 
vortices and charges (Cooper pairs), leading to
interesting phase transitions and dynamic responses in two dimensions.~\cite{Faziox91,BJKim95,GSS,qhe}
%In the vicinity of the superconductor-insulator transition,
%the fluctuation effects depend crucially
%on the dimensionality of the system.
%In the case of two-dimensional (2D) arrays,
%rich effects of quantum fluctuations and resulting phase transitions 
%have been examined rather extensively
%although there still exist unsettled issues
%in the quantum regime, such as low-temperature
%re-entrance~\cite{Faziox91,BJKim95}.
%On the other hand, one-dimensional (1D) chains of Josephson junctions, 
In one dimension,
where quantum fluctuations should be more important, 
the quantum phase transition has been studied 
in the absence of gate charges.~\cite{Bradle84} 
In such a one-dimensional (1D) array, i.e., a chain of Josephson junctions,
gate charges on grains, breaking the particle-hole symmetry,
can generate the persistent voltage in the appropriate regime~\cite{Choi93} 
and affect the quantum phase transition substantially,
which has been studied analytically via the mean-field approximation~\cite{MFA}
and via the perturbation expansion~\cite{pert}
as well as numerically via quantum Monte Carlo simulations.~\cite{MC}
In the analytical approaches, however, 
the second gives accurate results only in the weak-coupling limit,
i.e., for sufficiently small values of the Josephson coupling energy, whereas
the first is not expected to be reliable in one dimension, 
where fluctuations are too strong to be neglected. 
Both approaches fail to capture the essential physics of 
the Kosterlitz-Thouless-Berezinskii (KTB) transition,~\cite{Berezi71,Koster74}
present in the system without gate charges.
They not only disallow one to explore the detailed nature of the transition 
but also yield phase boundaries near the particle-hole symmetry line
in rather large discrepancy with those from simulations.~\cite{MC} 
%In particular the detailed nature of the transition for nonvanishing gate charges
%still remains unexplored.
It is thus desirable to study the gate charge effects 
in an analytical way beyond the above approaches, which discloses
the nature of the transition.

This paper presents the renormalization-group (RG) study of the 
quantum phase transition in a chain of Josephson junctions 
with gate charges.
At zero temperature, the system is mapped onto a two-dimensional (2D) classical
Coulomb gas, where the gate charge plays the role of an imaginary 
electric field.
This in general leads to the superconductor-insulator transition, 
driven by vortices,
as the ratio of the Josephson coupling energy to the charging energy is varied. 
Here the field associated with the gate charge is found to grow under the 
RG transformation and tend to suppress quantum fluctuations, 
helping to establish superconductivity in the system.
On the basis of this observation,
the zero-temperature phase diagram is proposed on the plane of
the gate charge and the energy ratio.

We consider a 1D array of $N$ superconducting grains, 
each coupled to its two nearest-neighboring grains via Josephson junctions
of strength $E_J$.
Each grain is characterized by the self-capacitance $C$, 
leading to the charging energy scale $E_C \equiv 4e^2 /C$,
and a gate charge $Q$ can be induced on it
externally, e.g., by applying gate voltage $V$ with respect to the
ground, giving $Q=CV$. 
Such a Josephson-junction chain is described by the Hamiltonian
\begin{equation}   \label{QPM}
%H = \frac{E_C}{2}\sum_{k=1}^N (n_k +q)^2
%    - E_J \sum_{k=1}^N \cos(\phi_k -\phi_{k+1})
H = \frac{1}{2K}\sum_{x=1}^N (n_x +q)^2
    - K \sum_{x=1}^N \cos(\phi_x -\phi_{x+1}) ,
\end{equation}
where the lattice constant has been set equal to unity,
the energy has been rescaled in units of the Josephson plasma
frequency $\hbar\omega_p \equiv \sqrt{E_C E_J}$,
the square root of the energy ratio
$K \equiv \sqrt{E_J /E_C}$ corresponds to the effective coupling, and
$q \equiv Q/2e$ is the uniform gate charge in units of the Cooper pair
charge $2e$.
The number $n_x$ of Cooper pairs and
the phase $\phi_x$ of the superconducting order parameter 
at site $x$ are quantum-mechanically conjugate variables: 
$[n_x, \phi_{x'}]=i\delta_{xx'}$.
Due to the periodicity and symmetry in $q$, we need to consider only the
range $0\leq q \leq 1/2$.

Following the standard procedure,~\cite{Bradle84,Choi93} we
introduce the imaginary time $\tau$ running on the interval
$[0,\beta]$, and write the partition function in
the imaginary-time path-integral representation:
\begin{equation}
Z
= \prod_{x,\tau}\sum_{n_{x,\tau}}
  \int_0^{2\pi}\frac{d\phi_{x,\tau}}{2\pi}\;
  \exp\left\{\sum_{\tau=0}^{\beta-1} A[n,\phi]\right\}
\end{equation}
with the action
%\onecolm
\begin{eqnarray}
A[n,\phi]
= &-&i\sum_{x}^N n_{x,\tau}\partial_\tau \phi_{x,\tau}
  -\frac{1}{2K}\sum_{x}^N (n_{x,\tau}+q)^2 \nonumber \\
  &+& K\sum_{x}^N \cos \partial_x\phi_{x,\tau}
  ,
  \label{QPM:L}
\end{eqnarray}
%\twocolm\noindent
where the inverse temperature has been rescaled according to
$\hbar\omega_p\beta\to\beta$, 
and the (imaginary) time slice $\delta\tau$ 
has been chosen to be unity (in units of $\hbar\omega_p$).
%~\cite{end_note:2}.
Here $\partial_x$ and $\partial_\tau$ denote 
the difference operator with respect to
the position $x$ and to the imaginary time $\tau$, respectively:
$\partial_x\phi_{x,\tau}\equiv \phi_{x+1,\tau}-\phi_{x,\tau}$ and
$\partial_\tau\phi_{x,\tau}\equiv \phi_{x,\tau+1}-\phi_{x,\tau}$,
and the zero-temperature limit $\beta\to\infty$ as well as the thermodynamic
limit $N\to\infty$ is to be taken.
We then apply the Villain approximation to
integrate out the phase variables $\{\phi_{x,\tau}\}$ and 
%
%to obtain the partition function in terms of an additional set of integer variables
%$\{m_{j,\tau}\}$ (as well as $\{n_{j,\tau}\}$).
%The corresponding action is given by
%\begin{equation}
%A[n,m]
%=  \frac{1}{2K_0}\sum_{i,j}n_{i,\tau}C^{-1}_{ij} n_{j,\tau}
%  + \frac{1}{2K_0}\sum_jm_{j,\tau}^2  ,
%  \label{L:1}
%\end{equation}
%where the two sets of integer variables satisfy the constraint
%$\partial_x m_{j,\tau}+\partial_\tau{}n_{j,\tau}=0$.
%This constraint is conveniently taken into account by introducing an integer
%field $A_{\tilde{j},\tilde\tau}$ defined on the space-time dual
%lattice $(\tilde{j},\tilde\tau)\equiv(j+1/2,\tau+1/2)$ in such a way that
%$m_{j,\tau}=-\partial_\tau{}A_{\tilde{j},\tilde\tau}$ and 
%$n_{j,\tau}=\partial_x A_{\tilde{j},\tilde\tau}$.
%Henceforth we will work on the dual lattice, and drop for simplicity
%the tilde sign over site indices.
%
perform the duality transformation~\cite{Josexx77} to write
the partition function 
%is thus written in terms of the unconstrained summation of 
in the form $Z=\sum_{\{s_{{\bf R}}\}} \exp\{- A[s]\}$
%\begin{equation}
%Z
%= \prod_{j,\tau}\sum_{A_{j,\tau}}
%  \exp\left\{-\sum_\tau L[A]\right\}
%  ,
%\end{equation}
with the action
%\onecolm
\begin{equation}
 A[s]
=  \frac{1}{2K}\sum_{{\bf R}}[(\partial_X s_{{\bf R}}+q)^2 +(\partial_Y s_{{\bf R}})^2],
\end{equation}
%\twocolm
where the integer field $s_{{\bf R}}$ is defined on the dual lattice site
${\bf R} \equiv (X, Y)$, dual to the original 2D space-time lattice 
${\bf r} \equiv (x, \tau)$.
The Poisson summation formula allows us to decompose
$s_{{\bf R}}$ into a real-valued field $\theta_{{\bf R}}$ and a new integer
field $m_{{\bf R}}$ describing the vorticity. 
%Since the boundary condition is not specified for the charge variables,
%we may shift 
Shifting the field variable in such a way that 
$\theta_{{\bf R}} - qX \equiv \tilde{\theta}_{{\bf R}}$,
we finally obtain
\begin{eqnarray}
Z&=& \prod_{{\bf R}} \sum_{m_{{\bf R}}} \int_{-\infty}^{\infty} d\theta_{{\bf R}}
   \exp\left[-\frac{1}{2K} \sum_{\langle {\bf R},{\bf R}'\rangle} 
           (\theta_{{\bf R}} -\theta_{{\bf R}'})^2 \right.\nonumber \\
          & &~~~~~~~+\left. 2\pi i\sum_{{\bf R}} m_{{\bf R}} (\theta_{{\bf R}} +qX) \right],
\end{eqnarray}
where the tilde sign has been dropped for simplicity.
This leads, apart from the spin-wave part, 
to the 2D system of classical vortices described by the action
\begin{eqnarray}  \label{2DVG}
A_c
&=& \pi K \sum_{{\bf R},{\bf R}'}
  m_{{\bf R}}\,G'({\bf R}-{\bf R}')\,m_{{\bf R}'} + 2\pi i q \sum_{{\bf R}} m_{{\bf R}} X
  \nonumber \\
&=& \pi K \sum_{{\bf R},{\bf R}'}
  m_{{\bf R}}\,\ln |{\bf R}-{\bf R}'|\,m_{{\bf R}'} + 2\pi i q \sum_{{\bf R}} m_{{\bf R}} X \nonumber \\
  & &~~+ \ln y \sum_{{\bf R}} m_{{\bf R}}^2  
\end{eqnarray}
with the vorticity neutrality condition $\sum_{\bf R} m_{\bf R} =0$,
where $G'({\bf R}-{\bf R}') \equiv 2\pi [G(0) - G({\bf R}-{\bf R}')]$
describes the lattice Coulomb Green's function with the singular
diagonal piece $G(0)$ subtracted
and $y$ corresponds to the vortex fugacity.
Note that the gate charge $q$ plays the role of an imaginary electric
field acting on vortices. 
Accordingly, 
the effects of the gate charge on the quantum phase transition 
in the Josephson-junction chain
reduces to those of the (imaginary) electric field in the 2D Coulomb gas.

To investigate the quantum phase transition displayed by
the action in Eq.~(\ref{2DVG}), we now consider the correlation function, 
with attention to the effects of the gate charge. 
To the lowest order in the fugacity $y$, the vortex correlation function
reads
\begin{equation} \label{vorcorr}
\langle m_0 m_{{\bf R}} \rangle = - 2y^2 R^{-2\pi K} \cos 2\pi qX,
\end{equation}
which, like the Hamiltonian in Eq.~(\ref{QPM}), depends periodically on $q$.
We also have the average vorticity
\begin{equation} \label{avvor}
\langle m_0 \rangle = - 2iy^2 \sum_{{\bf R}} R^{-2\pi K} \sin 2\pi qX = 0,
\end{equation}
upon summing over the orientation; the gate charge does not alter the 
average vorticity.
Renormalization of the phase correlation function by such vortices
then leads to the scaling equations describing the critical behavior.
The phase correlation function takes the form~\cite{Josexx77}
\begin{equation} \label{corr}
g({\bf r}{-}{\bf r}') \equiv \langle e^{i(\phi_{\bf r}-\phi_{{\bf r}'})}\rangle
 = g_{sw}({\bf r}{-}{\bf r}') g_v ({\bf r}{-}{\bf r}')
\end{equation}
with the spin wave part 
\begin{equation}
g_{sw}({\bf r}{-}{\bf r}') = \exp\left[-\frac{1}{2\pi K} G'({\bf r}{-}{\bf r}')\right]
\end{equation}
and the vortex contribution given by~\cite{Josexx77}
\begin{equation}
\ln g_v ({\bf r}{-}{\bf r}') = 
%\frac{1}{8} \sum_{{\bf R}_0, {\bf r}_0} \langle m_0 m_{{\bf r}_0} \rangle [({\bf r}_0 %\cdot \nabla_{{\bf R}_0})u_{{\bf R}_0}]^2 &=&
 -\frac{1}{4} y^2 \sum_{{\bf R}_0, {\bf r}_0} r_0^{-2\pi K} \cos 2\pi q x_0 
         [({\bf r}_0 \cdot \nabla_{{\bf R}_0})u_{{\bf R}_0}]^2 .
\end{equation}
Here Eqs.~(\ref{vorcorr}) and (\ref{avvor}) have been used, and the function
$u_{\bf R}$ 
% \equiv \sum_{{\bf R}'} G'({\bf R}{-}{\bf R}')[\eta_{{\bf R}'}^{\ell}-\eta_{{\bf R}'}^{r} ]$
defined on the dual lattice satisfies~\cite{comm1}
\begin{eqnarray}
\partial_X u_{\bf R} 
     &=& -\partial_Y [G'({\bf r}{-}{\bf R})-G'({\bf r}'{-}{\bf R})] \nonumber\\
\partial_Y u_{\bf R} 
     &=& \partial_X [G'({\bf r}{-}{\bf R})-G'({\bf r}'{-}{\bf R})].
\end{eqnarray}
Replacing the lattice summation over ${\bf r}_0$ 
by the integration in the polar coordinate $(r_0, \theta_0 )$
and noting 
$[({\bf r}_0 \cdot \nabla_{{\bf R}_0})u_{{\bf R}_0}]^2 
 = r_0^2 [\cos^2\theta_0 (\partial_{X_0} u_{{\bf R}_0})^2 
          + \sin^2\theta_0 (\partial_{Y_0} u_{{\bf R}_0})^2
          + 2\cos\theta_0 \sin\theta_0 
              (\partial_{X_0} u_{{\bf R}_0})(\partial_{Y_0} u_{{\bf R}_0}) ]$,
we obtain
\begin{eqnarray}
\ln g_v ({\bf r}{-}{\bf r}') &=& 
  -\frac{\pi}{4} y^2 \sum_{{\bf R}_0} \int_1^{\infty} dr_0 \,r_0^{3-2\pi K}\nonumber\\ 
   & &~~\times \left\{[J_0(2\pi qr_0)+J_2(2\pi qr_0)]
           (\partial_{X_0} u_{{\bf R}_0})^2 \right. \nonumber\\
   & &~~~~\left. + [J_0(2\pi qr_0)-J_2(2\pi qr_0)](\partial_{Y_0} u_{{\bf R}_0})^2]\right\} \nonumber \\
 &=& -2\pi^2 y^2 \int dr_0 \,r_0^{3-2\pi K} J_0(2\pi qr_0) G'({\bf r}{-}{\bf r}'),
\end{eqnarray}
where $J_0$ and $J_2$ are Bessel functions and we have used~\cite{comm}
\begin{eqnarray*}
\sum_{{\bf R}_0} [(\partial_{X_0} u_{{\bf R}_0})^2 + (\partial_{Y_0} u_{{\bf R}_0})^2]
&=& \sum_{{\bf R}_0}(\nabla_{{\bf R}_0} u_{{\bf R}_0})^2 = 4\pi G'({\bf r}{-}{\bf r}') \\
\sum_{{\bf R}_0} [(\partial_{X_0} u_{{\bf R}_0})^2 - (\partial_{Y_0} u_{{\bf R}_0})^2]
&=& 0.
\end{eqnarray*}

Equation (\ref{corr}) thus reduces to
\begin{eqnarray}
\ln g(r) &=& \ln g_{sw}(r) + \ln g_v (r) \nonumber \\
         &=& -\frac{1}{2\pi K} G'(r) - 2\pi^2 y^2 G'(r) 
              \int dr_0 r_0^{3-2\pi K} J_0 (2\pi qr_0 ) \nonumber \\
         &\equiv & -\frac{1}{2\pi K_{eff}} G'(r), 
\end{eqnarray} 
which defines the effective coupling $K_{eff}$.
It is then straightforward to derive the scaling equations
\begin{eqnarray} \label{RG0}
\frac{dK^{-1}}{d\ell} &=& 4\pi^3 y^2 J_0 (2\pi q) \nonumber \\
\frac{dy}{d\ell} &=& (2-\pi K)y \\
\frac{dq}{d\ell} &=& q , \nonumber
\end{eqnarray}
the third of which demonstrates that the gate charge is relevant, 
growing under the RG transformation: 
$q(\ell) = q e^{\ell}$
with $q \equiv q(\ell{=}0)$ being the initial (physical) value of the gate charge.
This allows us to write the scaling equations in the form
\begin{eqnarray} \label{RG1}
\frac{dK^{-1}}{d\ell} &=& 4\pi^3 y^2 J_0 (2\pi q e^{\ell}) \nonumber \\
\frac{dy}{d\ell} &=& (2-\pi K)y ,
\end{eqnarray}
which reduce, for $q=0$, to the standard scaling equations 
for the $XY$ model.~\cite{Koster74,Josexx77}
Accordingly, the system undergoes a KTB transition~\cite{Berezi71,Koster74,Josexx77}
from the insulating phase to the superconducting one as $K$ is
increased.~\cite{Bradle84}

When the gate charge is present ($q\neq 0$), on the other hand,
the Bessel function in Eq.~(\ref{RG1})
first decreases with $\ell$ and can even become negative as the renormalization
proceeds.
% ($\ell \rightarrow \infty$). 
The effective temperature $K^{-1}$ then increases less rapidly
and may even decrease under RG transformation; in this way the gate charge
tends to suppress quantum fluctuations, helping to establish superconductivity 
in the system.
Note also that the periodicity in $q$ has been lost in Eq.~(\ref{RG0}), which
results from the continuum approximation, i.e.,
the replacement of the lattice summation over ${\bf r}_0$ 
by the integration over $r_0$ and $\theta_0$. 
To be more accurate, we may avoid the continuum approximation and 
employ the Poisson summation formula:
\begin{eqnarray}
\sum_{{\bf r}_0} f({\bf r}_0)
&=& \sum_{j,k=-\infty}^{\infty} \int dr_0 r_0 \int d\theta_0  f(r_0, \theta_0)
  \nonumber\\
   & &~~~~\times \exp[2\pi ir_0 (j\cos\theta_0 + k\sin\theta_0)].
\end{eqnarray}
This leads to 
\begin{eqnarray} \label{RG}
\frac{dK^{-1}}{d\ell} = 2\pi^3 y^2 \sum_{j,k} 
  & & \left[ J_0 (2\pi\sqrt{(j{+}q)^2{+}k^2} e^{\ell}) \right. \nonumber\\
   & &~~~ \left. +J_0 (2\pi\sqrt{(j{-}q)^2{+}k^2} e^{\ell})\right] ,
\end{eqnarray}
which indeed manifests the periodicity,
in place of the first one in Eq.~(\ref{RG1}); the second one remains
unchanged.
For $q=0$, the terms of $j\neq 0$ as well as those of $k\neq 0$ 
give only small quantitative corrections,
and we may use the approximate scaling equations given by Eq.~(\ref{RG0}) or
(\ref{RG1}).~\cite{Koster74,Josexx77}
On the other hand, for $q \neq 0$, the contributions of nonzero 
$j$ terms can have crucial effects, and should be taken into account.

\begin{figure}
\begin{center}
\epsfig{file=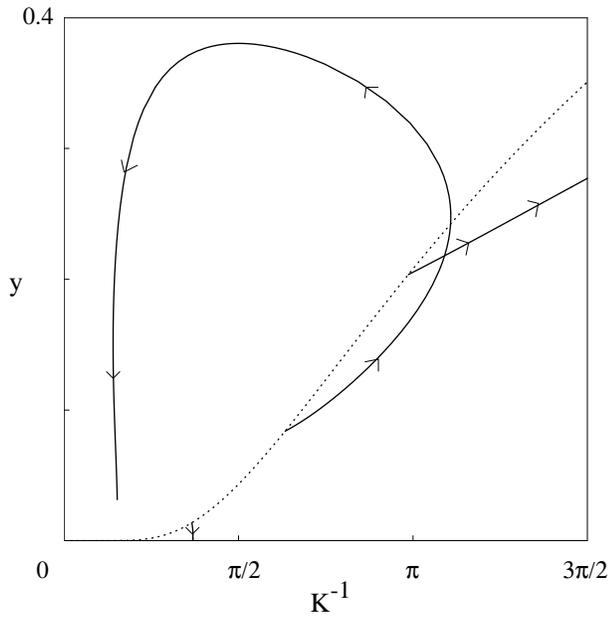,clip=,width=80mm} 
%\\
%\epsfig{file=JJc-fig1b.eps,clip=,width=80mm}
\end{center}
\caption{Typical renormalization-group flow diagram on the $K^{-1}{-}y$ plane.
Here the initial (physical) value of the gate charge is $q=0.1$ and the dotted line represents the locus of the initial configuration. 
}
\label{fig:RG}
\end{figure}

We have thus computed numerically the RG flow from the above scaling equations 
for various (initial) values of $q$ as well as $K$, and show
the typical RG flow in Fig.~\ref{fig:RG}.
The dotted line represents the initial configuration, i.e., the physical
value of the fugacity, chosen to be $y=e^{-(\pi^2/2)K}$.
It is obvious that the Gaussian fixed line ($y=0$) is stable for $K>2/\pi$ and
unstable for $K<2/\pi$.  When the initial coupling is strong $(K\gg 1)$, 
the system thus approaches the Gaussian fixed line with finite (renormalized) 
values of $K$, yielding superconductivity; at weak (initial) couplings $(K\ll 1$), 
the system flows to the high (effective) temperature limit 
($K^{-1}\rightarrow \infty$) and becomes insulating.
For intermediate couplings, the non-zero gate charge, growing under the
RG transformation, can make $K$ increase and alter the flow qualitatively:
The RG flow starting near $K\approx 2/\pi$ can eventually turn around, 
approaching the fixed line, as illustrated in Fig.~\ref{fig:RG}.
Note that the projections of the flow trajectories onto the $K^{-1}{-}y$ plane
are depicted here.  In fact the trajectories exist in the three-dimensional $(K^{-1},y,q)$
space, and do not cross with each other.
The minimum initial value of the coupling, which yields such turning 
flow, then gives the critical coupling strength $K_c$, below which the system
is driven by quantum fluctuations to the insulating state.
Via extensive numerical computation, we have obtained $K_c$ 
for various values of the gate charge $q$ and found that $K_c$,
starting from $0.748$ at $q=0$, decreases monotonically with $q$;
in this way the gate charge tends to suppress quantum fluctuations, 
helping to establish superconductivity in the system.
%Note also that the renormalized value $K_R$ approached by $K_c$ depends on
%$q$ and in general increases with $q$ from the universal value $2/\pi$
%at $q=0$.
Of particular interest is the case of the maximal gate charge ($q=1/2$),
where the presence of degeneracy is believed to restore superconductivity 
at arbitrarily weak couplings.~\cite{MFA,pert} 
The RG approach, with the contributions of several nonzero $j$ terms 
in Eq.~(\ref{RG}) taken into account, 
indeed indicates that $K_c$ reaches its
minimum at $q=1/2$.  Unfortunately, however, it is not allowed to confirm that 
$K_c$ actually vanishes in this case:
For large gate charges $(q\sim 1/2)$, the scaling equations yield
the RG flow going through the region
of large values of the fugacity $(y \gtrsim 1)$, 
for which the scaling equations themselves are not reliable. 

\begin{figure}
\begin{center}
\epsfig{file=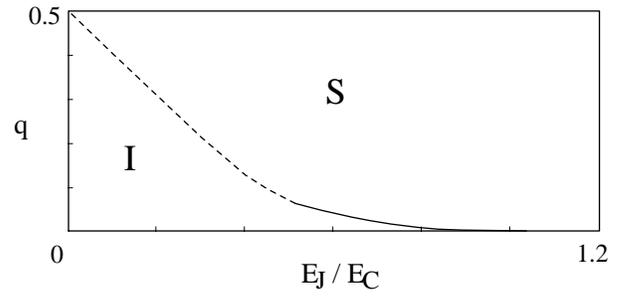,clip=,width=80mm} 
%\\  
%\epsfig{file=JJc-fig2b.eps,clip=,width=80mm} 
\end{center}
\caption{Phase diagram of a Josephson-junction chain, displaying the boundary between the
superconducting phase (S) and the insulating phase (I) on the $E_J/E_C{-}q$ plane.
The somewhat speculative boundary is depicted by the dashed line. 
}
\label{fig:ph_diag}
\end{figure}

The critical coupling strength $K_c$, obtained numerically from the
scaling equations as above, 
leads to the phase diagram in Fig.~\ref{fig:ph_diag}. 
Note that due to the employed Villain approximation, 
the coupling $K$ in the scaling equations
cannot be identified directly with $\sqrt{E_J/E_C}$ in the original Josephson-junction chain system.  
Comparison of the interaction form in the Villain action and the original one
shows that the critical coupling strength $K_c =0.748$ for $q=0$ 
corresponds to the critical value $(E_J/E_C)_c \approx 1.035$, 
which sets the overall scale. 
Accordingly, for each value of $q$, the obtained value of 
$K_c$ is transformed into the corresponding value of $(E_J/E_C)_c$, 
which gives the boundary between 
the superconducting phase and the insulating one in Fig.~\ref{fig:ph_diag}; 
manifested is the suppression of quantum fluctuations by the gate charge. 
At $q=0$ the transition between the two phases is of the standard KTB type, 
driven by vortices in the 2D space-time.  
On the other hand, the nonzero gate charge, which breaks the
particle-hole symmetry, makes a relevant perturbation and 
changes the transition in a significant way.
The critical RG flow corresponding to $K_c$ approaches a fixed point 
on the Gaussian fixed line with an arbitrarily large renormalized value 
instead of the standard KTB value $2/\pi$, thus 
suggesting a transition of the Gaussian nature. 
For large $q$, the RG flow goes through the region of large values
of the fugacity, for which the scaling equations are not reliable. 
The corresponding boundary, which is thus somewhat speculative, 
is drawn as a dashed line.
Note that the perturbation expansion,~\cite{pert} 
valid in the limit $E_J/E_C \rightarrow 0$, gives accurate results around
$q \approx 1/2$.  In this sense the RG approach here complements the
perturbation expansion. 
Here it is pleasing that Fig.~\ref{fig:ph_diag} agrees well with
the results of quantum Monte Carlo simulations,~\cite{MC}
particularly near the symmetry line, 
where the perturbation expansion as well as the mean-field approximation 
gives large discrepancy.

In summary, we have employed the renormalization-group transformation
method to investigate the quantum phase transition 
in a chain of Josephson junctions, with attention to the effects
of gate charges. 
At zero temperature the system has been mapped onto a two-dimensional 
classical Coulomb gas, which undergoes a superconductor-insulator 
transition as the energy ratio is varied. 
In the absence of the gate charge the transition, driven by vortices, 
is of the KTB type. 
Here the gate charge has been found to play the role of 
an imaginary electric field, which grows under the 
renormalization-group transformation. 
In particular, it not only changes the nature of the transition
but also tends to suppress quantum fluctuations, 
helping establish superconductivity in the system. 
On the basis of this observation,
we have obtained the zero-temperature phase diagram on the plane of
the gate charge and the energy ratio, which is, unlike the diagram
obtained in the previous analytical approaches, 
fully consistent with the results of quantum Monte Carlo simulations.

MYC and ML thank M.-S. Choi for useful discussions 
and D.J. Thouless for the hospitality during their stay at 
University of Washington, where this work was performed.
JC thanks D. Belitz for the hospitality during his stay at University
of Oregon. 
This work was supported in part by the NSF Grant DMR-9815932
and by the BK21 Program, Ministry of Education of Korea.

%\bibliographystyle{prsty}
%\bibliography{alias,jjc}

\begin{thebibliography}{10}
\bibitem{comm}
The second of the below involves a singular integral, which depends on the order of
integration on the $(x, \tau)$ plane.  However, the value can be fixed by imposing the
boundary condition that it should vanish in the limit ${\bf r}\rightarrow {\bf r}'$.
\end{thebibliography}

\narrowtext

\end{multicols}

\end{document}